\documentclass[doublespacing]{elsart}
\usepackage{graphicx}
\usepackage{amsmath}
\begin{document}
\begin{frontmatter}
\title{Electric and magnetic field optimization procedure for Penning trap mass spectrometers}
\author[gsi]{D. Beck},
\author[gsi,MPIK]{K. Blaum},
\author[MSU]{G. Bollen},
\author[CERN]{P. Delahaye},
\author[gsi]{S. George},
\author[lab1,dag]{C. Gu\'enaut},
\thanks[dag]{\textit{Present address:} Millennium, Courtaboeuf, France}
\author[gsi]{F. Herfurth},
\author[greis,cros]{A. Herlert}, \ead{alexander.herlert@cern.ch}
\thanks[cros]{\textit{Present address:} CERN, Physics Department, 1211 Geneva 23, Switzerland}
\author[lab1]{D. Lunney},
\author[greis]{L. Schweikhard},
\author[gsi]{C. Yazidjian}.
\address[gsi]{GSI, Planckstra\ss e 1, 64291 Darmstadt, Germany}
\address[MPIK]{Max-Planck-Institut f\"ur Kernphysik, Saupfercheckweg 1, 69117 Heidelberg, Germany}
\address[MSU]{NSCL, Michigan State University, East Lansing, MI 48824-1321, USA}
\address[CERN]{CERN, Physics Department, 1211 Geneva 23, Switzerland}
\address[lab1]{CSNSM-IN2P3-CNRS, 91405 Orsay-Campus, France}
\address[greis]{Institut f\"ur Physik, Ernst-Moritz-Arndt-Universit\"at, 17487 Greifswald, Germany}
\begin{abstract}
Significant systematic errors in high-precision Penning trap mass spectrometry 
can result from electric and magnetic field imperfections. An experimental procedure
to minimize these uncertainties is presented for the on-line Penning trap mass
spectrometer ISOLTRAP, located at ISOLDE/CERN. The deviations from the ideal magnetic and
electric fields are probed by measuring the cyclotron frequency and the reduced cyclotron
frequency, respectively, of stored ions as a function of the time between the ejection of
ions from the preparation trap and their capture in the precision trap, which influences
the energy of their axial motion.
The correction parameters are adjusted to minimize the frequency shifts.
\end{abstract}
\end{frontmatter}
\section{Introduction}
The mass of a nuclide is a fundamental property since it gives
access to the binding energy which reflects the net effect of all
forces at work in the nucleus\,\cite{Lun03}. It is of importance for
various fields such as atomic physics, chemistry, nuclear structure,
astrophysics, and the study of the weak interaction. Because the binding
energy is small compared to the overall atomic mass, the required
measurement accuracy is necessarily high.

Of the many techniques used for mass measurements, the Penning trap
has emerged as the instrument of choice for high
precision\,\cite{Bla06}, achieving a relative uncertainty of down to the
order of $10^{-11}$ on stable nuclides\,\cite{Rai04}, which allows
the probing of even the \textit{atomic} binding energy. For
radioactive species, ISOLTRAP\,\cite{Bol96,Muk08}, located at the
ISOLDE/CERN facility\,\cite{Kug00}, has been the pioneering Penning trap
experiment for on-line mass measurements of short-lived nuclei, meanwhile
routinely reaching a relative mass uncertainty
of $\delta m/m=10^{-8}$ \cite{Kel03}. ISOLTRAP has
continuously improved its accuracy and applicability.
Examples are the installation of a linear radiofrequency quadrupole
ion guide and beam buncher\,\cite{Her01}, the introduction of mass
spectrometry of ions produced by in-trap decay\,\cite{Her05}, the
systematic study of the measurement uncertainties using
carbon-cluster ions\,\cite{Kel03,Bla02,Bla03b}, the implementation
of a magnetron phase\,locking mechanism\,\cite{Bla03}, and the use
of Ramsey's technique for the excitation of the ion
motion\,\cite{Geo2007a,Geo2007b}.

ISOLTRAP has been followed by other Penning trap mass spectrometers,
which are in
operation or in preparation\,\cite{Bla06,Sch06}: SMILETRAP using
stable, highly charged ions \cite{Ber02}, the Canadian Penning
Trap\,\cite{Sav06}, JYFLTRAP\,\cite{Jok06}, SHIPTRAP\,\cite{Rah06},
LEBIT\,\cite{Rin06}, and TITAN\,\cite{Dil06}. New on-line traps are
now being commissioned: MLLTRAP at Munich\,\cite{Hab05} and
TRIGA-TRAP at the research reactor TRIGA Mainz\,\cite{Ket08}. 
For mass measurements in Penning traps, most of the systematic
errors arise from misalignment and magnetic- and electric-field
imperfections (see \cite{Bol96}). Therefore, all of the systems
mentioned above have to deal with similar problems concerning the
optimization of the electric and magnetic trapping fields.
In the case of ISOLTRAP, the mechanical misalignment was minimized
during the installation. The present work addresses the optimization
of the magnetic and electric fields.

\section{Ion motion in a Penning trap}\label{ion_motion}
In the following, the principles and properties of a Penning trap
will be briefly reviewed as far as they apply to high-precision mass
measurements. For a more detailed description of the theoretical
aspects of Penning traps, see \cite{Bro86,Bol90,Kre90}.

An ideal Penning trap is defined as the superposition of a
homogeneous magnetic field $B$ and an electrostatic quadrupole field
$V(\rho,z)$ coaxial to the magnetic field. The combination of these
particular fields allows to store charged particles in a
well-defined volume. Also, there is an exact solution of the
equations of motion in the case of a single stored ion.

\begin{figure}
\begin{center}
\includegraphics[width=11cm]{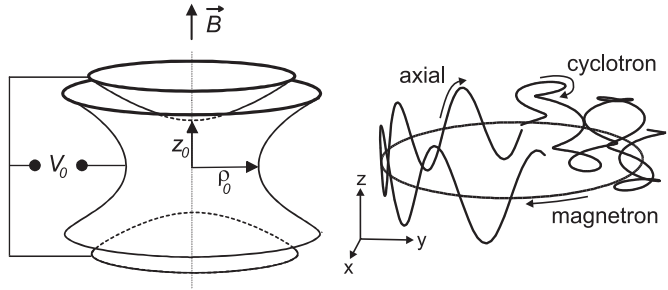}
\caption{\small{\textit{Left: Geometry of
a hyperboloidal Penning trap. Right: Schematic
representation of an ion trajectory, which is a superposition
of the three eigenmotions in a Penning trap.\label{fig:motion}}}}
\end{center}
\end{figure}

The electrostatic quadrupole field can be obtained by an electrode
configuration as shown in Fig.\,\ref{fig:motion}\,(left):
two endcaps and a ring electrode, all being hyperboloids of
revolution.

A potential difference $V_{0}$ (the trapping
potential) between the endcaps and the ring electrode creates the
quadrupolar potential

\begin{equation}
V(\rho,z)=\frac{V_{0}}{4d^{2}}(2z^{2}-\rho^{2}).
\end{equation}

The characteristic trap dimension $d$ is determined by

\begin{equation}
4d^{2}=(2z^{2}_{0}+\rho^{2}_{0}) \label{d},
\end{equation}

where $\rho_{0}$ is the inner radius of the ring electrode and
$2z_{0}$ the distance between the endcaps, as shown in
Fig.\,\ref{fig:motion}\,(left).

In a pure magnetic field the stored particle with mass $m$ and
charge $q$ performs a circular motion at the cyclotron frequency

\begin{equation}
\nu_{c}=\frac{q}{2\pi m}B.
\end{equation}

In the presence of the quadrupolar electrostatic field the ion
motion becomes a superposition of three independent harmonic motions
as illustrated in Fig.\,\ref{fig:motion}\,(right). The ions have an
axial oscillation mode with frequency

\begin{equation}
\nu_{z}=\frac{1}{2\pi}\sqrt{\frac{qV_{0}}{md^{2}}}
\end{equation}

and two circular radial modes, the cyclotron and the
magnetron motion with eigenfrequencies $\nu_{+}$ (reduced cyclotron frequency)
and $\nu_{-}$ (magnetron frequency), respectively, given by

\begin{equation}\label{eq_om+-}
\nu_{\pm}=\frac{\nu_{c}}{2}\pm
\sqrt{\frac{\nu^{2}_{c}}{4}-\frac{\nu^{2}_{z}}{2}}.
\end{equation}

The sum of the two radial eigenfrequencies obeys the relation:

\begin{equation}\label{eq_omc}
\nu_{c}=\nu_{+}+\nu_{-}.
\end{equation}

A direct excitation of the ion motion at this sum frequency with an azimuthally
quadrupolar rf field \cite{Bol90} allows a mass
determination of the stored ion which relies only on the magnetic
field $B$. At the same time $\nu_c$ is a sensitive probe of the magnetic field
strength experienced by the ions and will be used for the magnetic field
optimization. In contrast, for the optimization of the electric field the
reduced cyclotron frequency $\nu_+$ will be investigated.

Another important relation between the eigenfrequencies
with respect to mass spectrometry is the so called
\emph{Invariance Theorem}\,\cite{Bro82}:

\begin{equation}\label{eq_omc-2}
\nu_{c}^2=\nu_{+}^2+\nu_{-}^2+\nu_z^2,
\end{equation}

which is independent of field imperfections to first order. However,
since only Eq.~(\ref{eq_omc}) is applied in high-precision Penning
trap mass measurements on short-lived nuclides, the largest source
of uncertainties are electric and magnetic field imperfections that
cause a broadening and a shift of the cyclotron frequency resonance.
These imperfections are due to the fact that a real Penning trap
deviates from the ideal case in many aspects \cite{Bol96,Bro86}. In
the following, the electric and magnetic field imperfections and their
minimization are discussed using the ISOLTRAP mass spectrometer as
an example.

\begin{figure}
\begin{center}
\includegraphics[width=8.5cm]{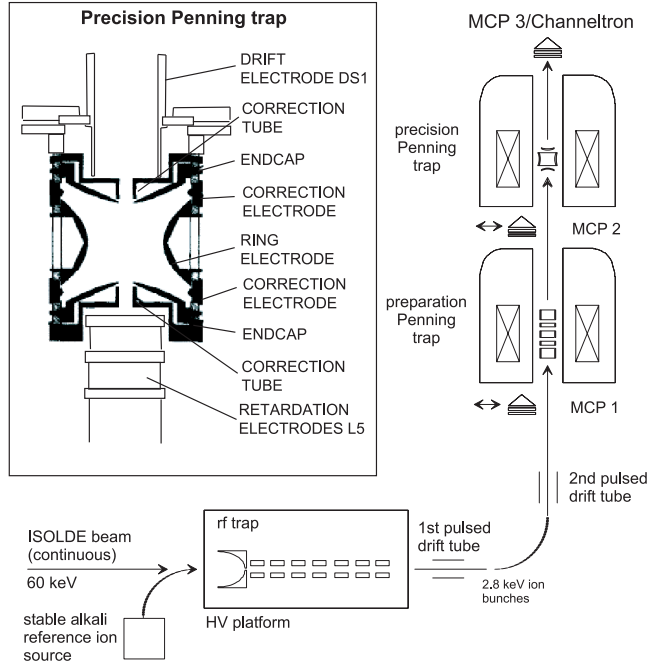}
\caption{\small{\textit{Experimental setup of the ISOLTRAP mass
spectrometer with the three main parts: A gas-filled linear
radio-frequency quadrupole (RFQ) trap, a gas-filled cylindrical
Penning trap, and a high-vacuum hyperboloidal Penning trap. For the
time-of-flight measurement a Channeltron detector is
installed 1.2\,m upstream \cite{Yaz06}.
The inset shows a more detailed illustration
of the precision Penning trap.\label{fig:ISOLTRAP}}}}
\end{center}
\end{figure}

\section{Experimental setup}
\begin{figure}
\begin{center}
\includegraphics[width=8.5cm]{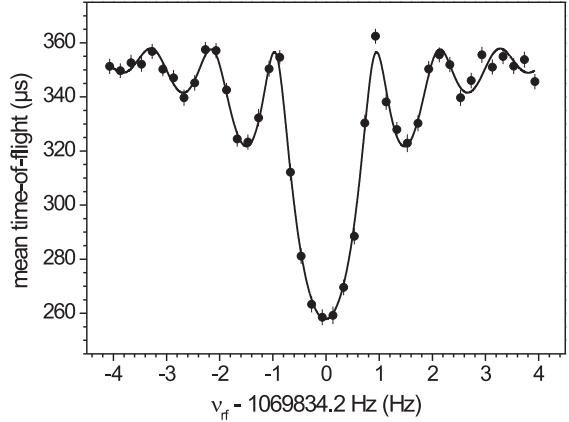}
\caption{\small{\textit{Cyclotron resonance of $^{85}$Rb$^+$ ions recorded
for an excitation duration of 900\,ms. The solid line is a fit
of the line shape to the data points.\label{fig:resonance}}}}
\end{center}
\end{figure}

ISOLTRAP is a high-precision Penning trap mass spectrometer
consisting of three main parts (see
Fig.\,\ref{fig:ISOLTRAP})\,\cite{Muk08}: First, a gas-filled
radio-frequency quadrupole (RFQ) trap serves as a cooler and buncher
to adapt the 60-keV ISOLDE ion beam to the ISOLTRAP requirements
with respect to kinetic energy, time structure, and beam
emittance\,\cite{Her01}. Second, a high-capacity cylindrical Penning
trap is used for isobaric cleaning of the beam by exploiting a
mass-selective helium-buffer gas cooling technique\,\cite{Sav91}
with a resolving power of up to $10^5$\,\cite{Rai97}. Finally, the
cooled ion bunch is transferred to the precision Penning trap where
the mass measurement is carried out by the determination of the
cyclotron frequency of the stored ions, using a time-of-flight (ToF)
detection method\,\cite{Gra80}: After a dipolar radiofrequency (rf)
excitation of the ions to a magnetron orbit of
about 0.7-1.0\,mm radius \cite{Bla03},
the initially pure magnetron motion is converted into cyclotron
motion by a quadrupolar rf field \cite{Konig}. At $\nu_{rf}=\nu_{c}$
a full conversion from initially pure magnetron to pure cyclotron
motion is obtained. In this case, the orbital magnetic moment $\mu$ and
the radial kinetic energy $E=\mu B$ are increased. The ions are
ejected from the trap and their ToF to an ion detector is measured.
Since there is an axial
acceleration of the ions in the fringe field of the superconducting
magnet, which is proportional to $\mu$, the shortest ToF is observed for
$\nu_{rf}=\nu_{c}$. Figure\,\ref{fig:resonance} shows a typical resonance
where the ToF is measured as a function of the frequency $\nu_{rf}$ 
of the excitation field applied.
The mass of the ion of interest is obtained from
the comparison of its cyclotron frequency with that of a well-known
"reference mass", provided from either ISOLDE or an off-line
reference ion source. The measurement procedure as well as the study
of the accuracy limit and systematic uncertainties are described in
detail in\,\cite{Kel03}.

\subsection{Magnetic field imperfections}
In order to perform a high-precision cyclotron-frequency
measurement, excellent homogeneity and temporal stability of the
magnetic field are required. Here, the possibility to minimize
magnetic field inhomogeneities by a dedicated optimization procedure
is addressed. The temporal stability is discussed in
\cite{Kel03,MMJ07}.

Magnetic field imperfections arise from the homogeneity limits of
commercial superconducting magnets: nowadays an inhomogeneity of
typically $\Delta B/B < 10^{-7}$ can be provided over a volume of
about a cubic centimeter. The volume typically probed by the ions
during precision mass measurements
results from the amplitudes $a_\pm\approx1$\,mm and $a_z\leq1$\,mm of
the radial and axial motional modes, respectively.
During the optimization the axial amplitudes are varied and increased
on purpose by up to an order of magnitide (see below).
For comparison, the precision trap has
dimensions $\rho_0=13.00$\,mm and $z_0=11.18$\,mm.
In addition to the intrinsic inhomogeneity of
the superconducting magnet, the homogeneity can
be disturbed if materials (including trap components) with a
magnetic susceptibility are introduced into the magnetic field. To
minimize this problem, the ISOLTRAP electrodes are made from
oxygen-free copper and the amount of material closest to the trap
center was minimized.

There are higher order magnetic field components that must be
minimized. Due to the mirror symmetry (with respect to the
$xy$-plane through the trap center) only even components occur. 
In addition, odd terms do not matter since they do
not result in frequency shifts, assuming that the average center of
the ion motion does not change. The frequency shift caused by
magnetic field inhomogeneities can be approximated by\,\cite{Bro86}

\begin{equation}
\Delta\nu_c^{magn}\approx \beta_2\nu_c(a^2_z-a^2_-/2),
\end{equation}

where $\beta_2$ denotes the relative strength of the lowest-order
component of magnetic inhomogeneities.

\begin{figure}
\begin{center}
\includegraphics[width=10cm]{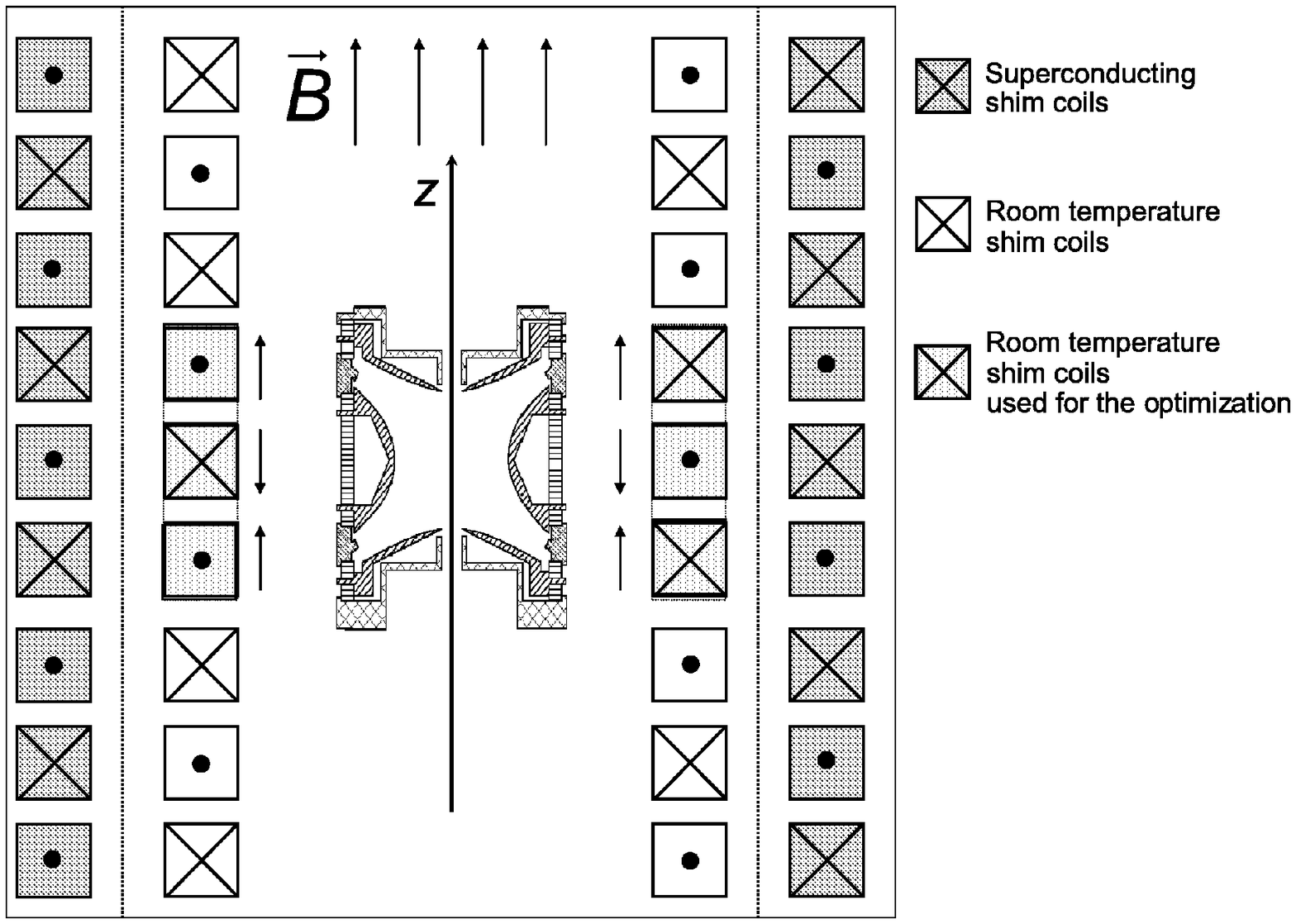}%
\caption{\small{\textit{A section of the measurement trap, which is
installed in a superconducting magnet (lateral cut) that generates a magnetic
field $B$ in the axial direction.
Several superconducting shim coils and room-temperature shim coils
are shown. The correction for magnetic field inhomogeneities is done
using the latter. Points and crosses represent the current direction
out of and into the plane, respectively.}}}\label{fig:magnet}
\end{center}
\end{figure}

The precision Penning trap (see inset of Fig.\,\ref{fig:ISOLTRAP})
is placed in a 5.9-T magnetic field generated by a superconducting
solenoid. To provide a homogeneous magnetic field the magnet is
equipped with a set of superconducting shim coils and a set of
room-temperature shim coils (see Fig.\,\ref{fig:magnet}). The
current on all superconducting shim coils was optimized during the
installation of the magnet. For practical use, only the
room-temperature shim coils are optimized routinely.
The three coils closest to the trap (dashed boxes in
Fig.\,\ref{fig:magnet}) have the strongest influence on the field
distribution and thus on the field homogeneity along the $z$-axis
inside the trap. Consequently, the optimization parameter is the
current $I_{S}$ applied simultaneously to these three
room-temperature shim coils. This current induces a magnetic field
oriented in one direction for the two outer coils and in the
opposite direction for the inner one (see Fig.\,\ref{fig:magnet}),
thus allowing to shim the $z^2$-component of the magnetic field, which can
be approximated along the axial direction by

\begin{equation}
B(z)=B_0(1+b_1z^2+b_2z^4+\dots).
\end{equation} 

Only the $z^2$-component is addressed since the axial amplitudes
are in general much larger than the radial ones and odd-terms cancel out.

\subsection{Electric field imperfections}
Electric field imperfections arise from deviations of the mechanical
trap construction from the ideal hyperbolical shape, 
such as holes in the endcaps for injection or
ejection of ions and the unavoidable truncation of the
electrodes. Thus, the electric field inside the trap does not follow
the pure quadrupolar form. However, it needs to be as ideal as
possible in order to assure the exact condition given in Eq.\,(\ref{eq_omc}).

For a real Penning trap the potential inside the trap can be expanded
in the form \cite{Bol90}:

\begin{eqnarray}\label{eq_VC}
V(\rho,z)&=&\frac{1}{2}V_{0}\Bigg[\frac{C_{2}}{d^{2}}\Big(z^{2}-\frac{1}{2}\rho^{2}\Big)+\frac{C_{4}}{d^{4}}
\Big(z^{4}-3z^{2}\rho^{2}+\frac{3}{8}\rho^{4}\Big)
\nonumber\\
&+&\frac{C_6}{d_6}\Big(z^{6}-\frac{15}{2}z^4\rho^2+\frac{45}{8}z^2\rho^4-\frac{5}{16}\rho^6\Big)+...\Bigg].
\end{eqnarray}

with $C_2$ as the quadrupole component. In an ideal trap $C_2=1$ and
$C_n=0$ for $n>2$. The frequency shift
$\Delta\nu_{c}^{elec}$\,\cite{Bol96}, which is due to the octupole
(represented by $C_4$) and dodecapole ($C_6$) contributions, depends
on the ion-motion amplitudes $a_{+}$, $a_{-}$, $a_{z}$ of the
cyclotron, magnetron, and axial motion, respectively, and is
given by

\begin{eqnarray}\label{eq_VD}
\Delta\nu_{c}^{elec}&=&\frac{V_0}{4\pi d^2B}\Bigg[\frac{3}{2}\frac{C_4}{d^2}(a^2_--a^2_+)\nonumber\\
&+&\frac{15}{4}\frac{C_6}{d^4}\big(a^2_z(a^2_--a^2_+)-(a^4_--a^4_+)\big)\Bigg].
\end{eqnarray}

To minimize the imperfections, additional electrodes are
implemented in the precision Penning trap: two correction rings and
two correction tubes, as shown in Fig.\,\ref{fig:ISOLTRAP}, to
eliminate $C_4$ and $C_6$. Without these additional electrodes
the multipole contributions to the potential are $C_2=0.96$, 
$C_4=0.23$, and $C_6=-0.26$ as deduced from simulations. The
correction ring electrodes lead to contributions of $C_4=-5.5\times10^{-4}$
and $C_6=1.5\times10^{-4}$ and have only an influence on the $C_2$-value
of the order of $1.6\times10^{-5}$. Similarly, the correction tube electrodes
lower the multipole contributions to $C_4=5.6\times10^{-3}$
and $C_6=9.0\times10^{-3}$ and have an effect on $C_2$ of the order of
$5.4\times10^{-3}$.
The remaining deviations from a
pure quadrupolar field are far away from the ion trajectory,
since for typical precision mass measurements
the amplitude of the axial motion does not exceed 1\,mm. 

\section{Field-optimization results}

\subsection{Probing the field imperfections}
The electric field optimization consists of varying the voltages of
the correction rings and correction tubes and monitoring the effect
on the reduced cyclotron frequency while the trapping conditions are varied.
After the buffer-gas cooling in the first Penning trap the axial
energy of the ions is thermalized to room temperature. From there
the ions are ejected and accelerated towards the precision Penning
trap. They are captured by, first, retarding them in the trap center
while the potential of the lower endcap has been lowered and,
second, restoring the potential of the lower endcap to its original
value. The capture time $T_{\text{cap}}$, a key parameter in the
field-optimization procedure, is defined as the time
between the ejection from the preparation Penning trap and the
rising of the lower endcap potential in the precision trap.

\begin{figure}
\begin{center}
\includegraphics[width=10cm]{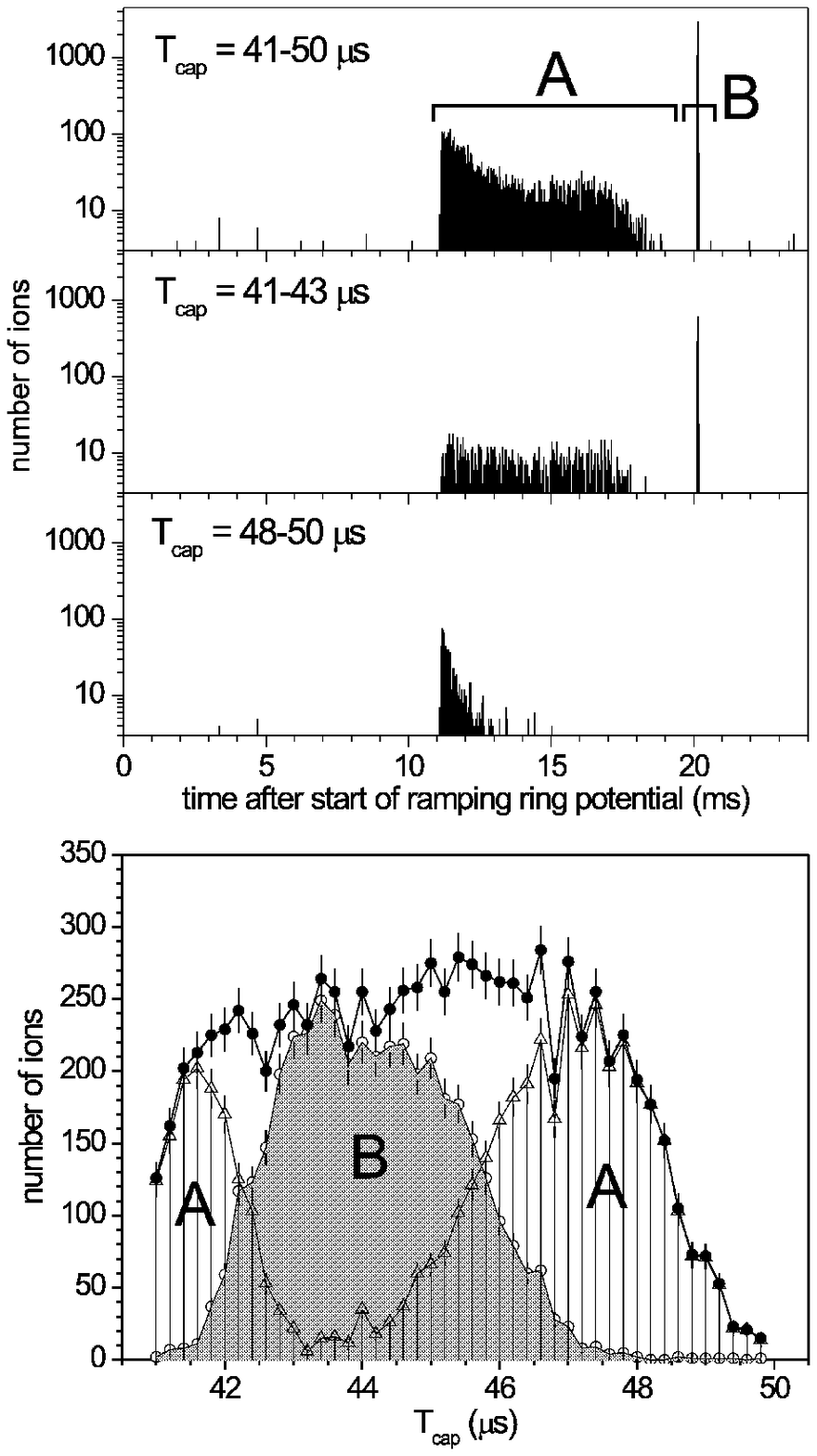}
\caption{\small{\textit{
Top: Time-of-flight spectra of $^{85}$Rb$^+$ ions being released
from the precision Penning trap towards the detector for different
ranges of the capture time $T_{\text{cap}}$ (see text for details).
Bottom: Number of captured ions as a function of the
capture time for $^{85}$Rb$^+$. The full symbols correspond to all
ions stored after ion capture. The open triangles denote ions which are
released during the ramping of the ring potnetial (range "A" in the ToF spectrum).
The open circles (curve filled with
a gray area) correspond to the number of ions that remain in the
trap after lowering the capture potential depth (range "B" in the ToF spectrum).
\label{hotcoldions}}}}
\end{center}
\end{figure}

Depending on both the initial axial energy distribution of the ions and
the capture time in the precision trap, the stored ions have
different axial energies and thus exhibit trajectories which may
cover a smaller or larger volume in the trap. Therefore, the
magnetic field and the quadrupolar electric field in the precision
trap should have as little as possible deviation from the ideal
fields in order to give the same conditions irrespective of the
initial axial energy after capturing. By the same token, the
resonance frequencies can be measured as a function of the capture time
in order to probe the field imperfections.

In general, before ejection towards the last ion detector (see
Fig.\,\ref{fig:ISOLTRAP}), the trapping potential depth is lowered
by ramping of the potential of the ring electrode (see \cite{Yaz06}
for details). In order to demonstrate the influence of the capture
time on the axial energy of the ions, the release of ions from the
trap during this ramping and at the moment of pulsing the endcap
is monitored as a function of the capture time. 
Fig.\,\ref{hotcoldions}\,(top) shows three ToF spectra accumulated
for different ranges of the capture time $T_{\text{cap}}$, where
the time $t=0$ marks the start of the ramping of the ring elecrode
potential from $-10$\,V to $-2.5$\,V, which lasts 17.5\,ms.

Ions with a higher axial energy than the trapping potential depth are
released and may reach the detector producing a signal in the time
range "A". After 19.5\,ms the endcap potential is pulsed and all ions
still stored in the trap are ejected. These ions produce the signal
in the time range "B". While stored in the trap, they
have low axial energies and thus low axial amplitude. In the normal
operation mode for precision mass measurements, only the corresponding
capture time is applied. For the present study, however, the capture time
will be varied systematically to observe its effect on the resonance frequencies.

Fig.\,\ref{hotcoldions}\,(bottom) gives the number of ions before and
during the ramping of the ring electrode (open triangles)
and after the ramping (open circles) as a function of the capture time.
The full circles correspond to the sum of all ions stored in the precision
trap after capturing. The axial energy of the stored ions after
capturing is mainly influenced by the initial energy of the ions as
they leave the preparation trap: ions with low axial kinetic energy
arrive at a later time than ions with higher axial kinetic energy
relative to the capture time in the precision trap. In addition,
ions with high axial kinetic energy may be reflected back to the
trap center by the upper endcap potential. Altogether, an asymmetric
distribution of ions with high and low axial kinetic energy is
obtained upon variation of the capture time.

\begin{figure}
\begin{center}
\includegraphics[width=9cm]{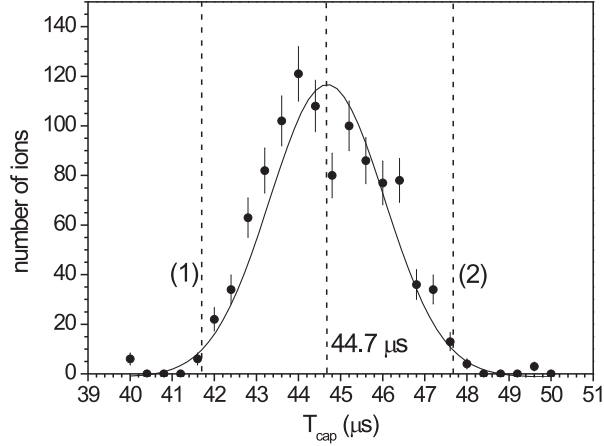}
\caption{\small{\textit{Number of captured ions as a function of the
capture time for $^{85}$Rb$^+$. The solid line is a Gaussian fit to
the data points. The FWHM of the fit is 3.5\,$\mu$s. The dashed lines
(marked "(1)" and "(2)") indicate the mimimum and maximum capture times
as used in the present investigation.
\label{delay}}}}
\end{center}
\end{figure}

In order to investigate the field corrections, the capture time
$T_{\text{cap}}$ is varied around the optimum value as shown in
Fig.\,\ref{delay} for $^{85}$Rb$^+$ ions. For non-optimal capture
times, the axial energy of the ions will be larger and deviations
from the ideal electromagnetic fields can be probed by measuring the
change of the ion-motion frequencies as compared to the case of the
optimal capture time. Since $\nu_c$ is directly related to $B$ and
not to $V_0$, we measure $\nu_c$ to probe the magnetic field and
$\nu_+$ (related to $V_0$) for the electric field optimization.

\subsection{Magnetic field}
The trim parameter for the magnetic-field homogeneity along
the $z$-axis is the current $I_{S}$ applied to the three
room-temperature shim coils close to the trap (see
Fig.\,\ref{fig:magnet}). The best value of the shim-coil current
will optimize the magnetic field homogeneity over the large axial
distance the ions cover, hence minimizing frequency variations for
deviating capture times and axial ion energies. To check the
magnetic field homogeneity, the cyclotron frequency $\nu_c$ is taken as
a probe, since it directly depends on the magnetic field (see
Eq.\,(\ref{eq_omc})). The cyclotron frequency is determined via the
measurement of time-of-flight resonance curves (see above) and
frequency changes are probed for different capture times
$T_{\text{cap}}$, i.e. in the present case between 41.7\,$\mu$s (1)
and 47.7\,$\mu$s (2) for $^{85}$Rb$^+$ (see Fig.\,\ref{delay}), and
different shim-coil currents.

\begin{figure}
\begin{center}
\includegraphics[width=9cm]{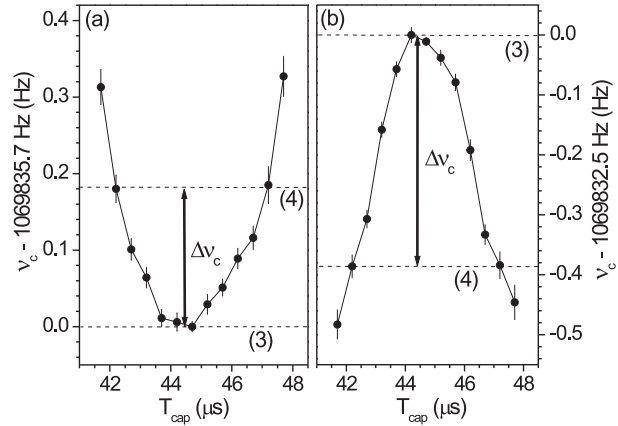}
\caption{\small{\textit{\label{fig:both_B}Cyclotron frequency $\nu_{c}$
as a function of the capture time $T_{\text{cap}}$ for two different
shim-coil currents: (a) $I_{S}=350$\,mA and (b) $I_{S}=100$\,mA.}}}
\end{center}
\end{figure}

The importance of such an optimization is shown in
Fig.\,\ref{fig:both_B}, which gives the cyclotron frequency as a
function of the capture time for two extreme values of the shim-coil
current. The corresponding relative frequency deviation for
$\Delta T_{cap}=\pm2.5\,\mu$s for each current setting is $\Delta
\nu_{c}/\nu_{c}= 4\cdot10^{-7}$. Note that there is a shift in
absolute frequency as well as an inflection in the frequency
variations. This is due to the fact that the correction coil design
was made such that a $B_0$ contribution is minimized and the
contribution to the next higher order term $B_2$ maximized. The
shift is a small left-over because the cancelation of $B_0$ from the
outer and the inner coil is not perfect.

\begin{figure}
\begin{center}
\includegraphics[width=9cm]{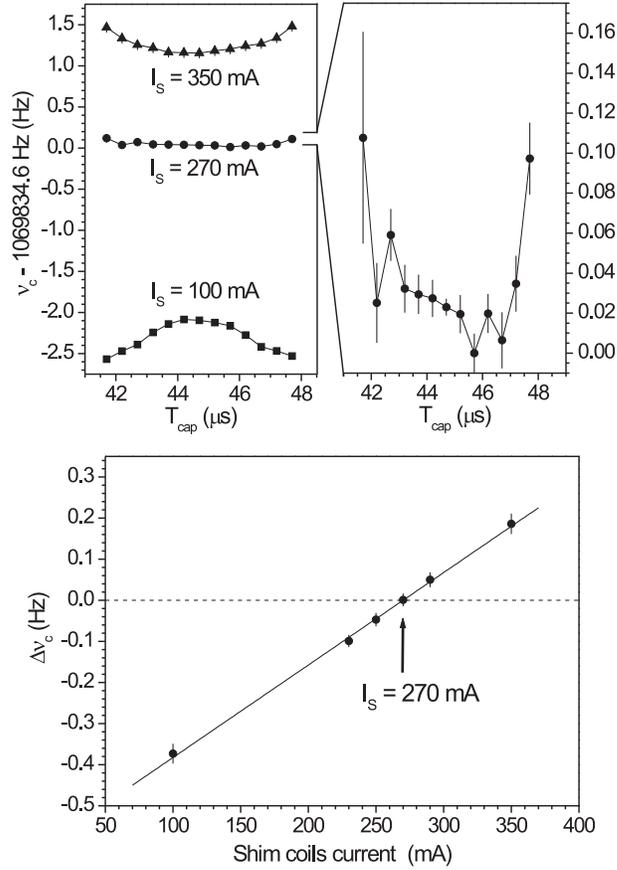}
\caption{\small{\textit{\label{fig:Opt_BAll_B}
Top: Cyclotron frequency $\nu_c$ as a function of the variation of
$T_{\text{cap}}$ for two extreme values of the shim-coil current
(full triangles: $I_{S}=350$\,mA; full squares: $I_{S}=100$\,mA) and for
the interpolated optimum value $I_{S}=270$\,mA (full circles). 
The right-hand side plot gives a zoom
for the optimum shim-coil current of $I_{S}=270$\,mA.
Bottom: Cyclotron frequency
difference $\Delta\nu_{c}$ between two capture times,
$T_{\text{cap}}=44.7\mu$s and 47.2\,$\mu$s, as a function of the
shim-coil current. The linear fit to the data points provides the
optimal value, i.e. where $\Delta \nu_{c}=0$: $I_{S}=270\pm3$\,mA.}}}
\end{center}
\end{figure}

To find the optimum value for $T_{\text{cap}}$, i.e. for which
$\Delta \nu_{c}=0$, the shim-coil current is varied and the
frequency shift $\Delta \nu_{c}$ between two different capture times
(marked as (3) with $T_{\text{cap}}$\,=\,44.7\,$\mu$s and (4) with
$T_{\text{cap}}$\,=\,47.2\,$\mu$s in Fig.\,\ref{fig:both_B}) is
measured. A capture time of 47.7 $\mu$s was not used because of the
low capturing efficiency for this value and thus the resulting long
measurement time. Moreover, it is expected that during a "normal" mass
measurement ions with the respective amount of axial energy will not
be present in the trap. The frequency variation was
measured for several shim-coil currents and the results are plotted
in Fig.\,\ref{fig:Opt_BAll_B}. The optimum value was deduced from a
linear fit to the data points, yielding $I_{S}=270$\,mA.

Cyclotron frequencies $\nu_c$ obtained with this optimum value
are presented in Fig.\,\ref{fig:Opt_BAll_B}\,(right). The relative deviation is
well below $5\cdot10^{-8}$ for a capture time variation of
$\pm$\,1.5\,$\mu$s. For a typical uncertainty in the capture-time
setting of about $\pm$\,0.3\,$\mu$s the relative frequency shift is
well below $1\cdot10^{-8}$. Figure\,\ref{fig:Opt_BAll_B}\,(left) also illustrates
the overall behavior of the cyclotron frequency as a function of the
applied shim-coil current, where an offset of the absolute frequency
and a relative frequency shift can be observed as a function of $T_{\text{cap}}$.
The latter lead to systematic errors thus pointing out the importance
of the magnetic field optimization and on making sure that the ions
are trapped in the center on average.

\subsection{Electric field}
To minimize electric-field
imperfections, two correction rings and tubes are used as shown
in Fig.\,\ref{fig:ISOLTRAP}. The correction rings correct
for the finite dimension of the electrodes while the correction
tubes compensate the discontinuity in the endcap surface due to the
ion entrance and exit holes. The influence of the rings is by far
much smaller than the influence of the tubes (see section\,\ref{rings}).

\subsubsection{Correction tubes}

\begin{figure}
\begin{center}
\includegraphics[width=9cm]{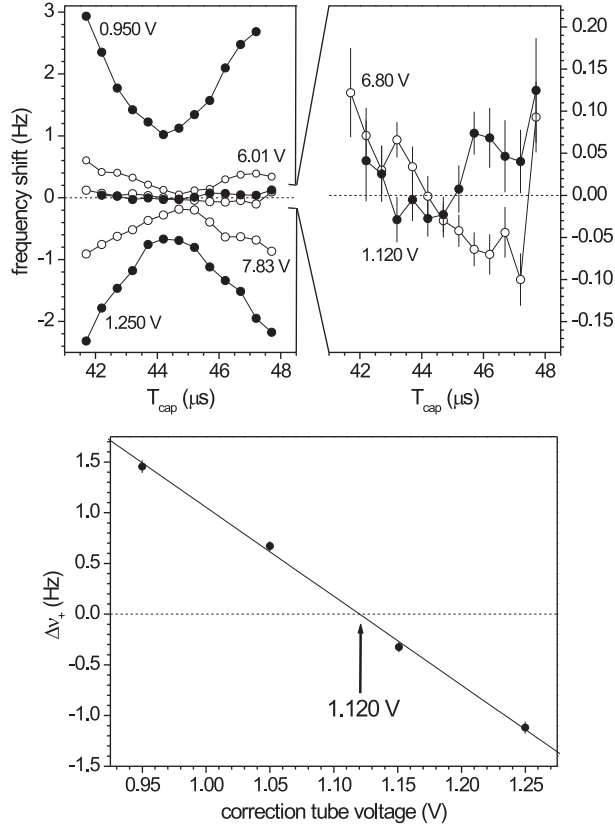}
\caption{\small{\textit{\label{fig:Both_E}
Top: Reduced cyclotron frequency as a function of
the variation of the capture time for different values of the
voltages on the correction tubes (0.950\,V, 1.120\,V, and 1.250\,V;
full circles) and rings (6.01\,V, 6.80\,V, and 7.83\,V; empty
circles). The zoom shows the frequency shift for the optimum values of
the correction tubes and rings voltage.
Bottom: Frequency shift
between two capture times ($T_{\text{cap}}$ = 44.7 $\mu$s and 47.2
$\mu$s) as a function of the correction tubes voltage. The linear
interpolation provides the optimal value: $(1.120\pm0.003)$\,V for
$\Delta\nu_{+}=0$.}}}
\end{center}
\end{figure}

The two correction tubes are called "Lower Correction Tube" (LCT),
and "Upper Correction Tube" (UCT). In general, the same voltage is
applied to both tubes. This voltage is the parameter used for the
optimization of the electric field. Again, the capture time is used
to probe the effects of the detuned field on the cyclotron
frequency.

The optimum value of the correction-tube voltage is found when an
ideal electric field is realized in the trap and therefore no
frequency shifts occur for all capture times
between 41.7\,$\mu$s and 47.7\,$\mu$s (see Fig.\,\ref{delay}). For
the electric field optimization, the reduced cyclotron frequency $\nu_{+}$
is monitored (see Eq.\,(\ref{eq_om+-})). The magnetic field optimization was
performed first to ensure that $B$ and thus $\nu_{c}$ are constant along
the axial direction of the trap. Consequently the variation of
$\nu_{+}$ is directly related to the electric field via $V_{0}$ with
the approximation $\nu_{+}=\nu_{c}-V_{0}/4\pi Bd^{2}$.

Figure\,\ref{fig:Both_E}\,(top) presents a measurement of the reduced
cyclotron frequency as a function of the capture time for different
values of the correction-tube and -ring voltage. The observed
curvatures with large relative frequency deviations $\Delta\nu/\nu$
show the importance of the electric-field optimization.

To determine the optimum value, the frequency difference
$\Delta\nu_{+}$ between two different capture times
($T_{\text{cap}}=44.7\,\mu$s and $T_{\text{cap}}=47.2\,\mu$s)
is measured as a function of the correction-tube voltage. The
results are plotted in Fig.\,\ref{fig:Both_E}\,(bottom). The optimum value
with a minimal frequency shift $\Delta\nu_{+}$ was found by
interpolation to be 1.120\,V . Results with this optimum value are
presented in Fig.\,\ref{fig:Both_E}\,(top). Again, in addition to the
frequency variations caused by the variation of the capture time,
shifts of the frequency at the optimal time are observed for
different voltages. After the optimization the deviation $\Delta
\nu_{+}/\nu_{+}$ is minimized to well below $1\cdot10^{-8}$ for
the capture-time range $T_{\text{cap}}=(44.7\pm1.0)\,\mu$s. The
remaining deviations may be due to temperature fluctuations in the
ISOLDE hall \cite{MMJ07}.

\subsubsection{Correction rings} \label{rings}
Figure\,\ref{fig:Both_E}\,(top) shows a comparison of the influence of
the correction rings and the correction tubes. For this study the
optimal value for the voltage applied to the correction rings and to
the correction tubes (i.e. to obtain a flat dependence as a
function of $T_{\text{cap}}$) was used and also two extreme values
chosen in the same proportion for the rings and the tubes. For the
measurements of the influence of the correction rings the correction
tubes were set to the optimum value of 1.120\,V. Vice versa during
the measurement of the influence of the correction tubes, the
correction rings were set to the optimum value of 6.80\,V.

The influence of the correction rings is about five times smaller
than that of the tubes, so the optimization should be focused on the
tubes. This was expected from the simulations as mentioned above,
which show a stronger influence of the tubes. The
difference between the behavior observed in other
systems\,\cite{Ger90} and the ISOLTRAP result is due to the fact that the
axial motion of the ions is not cooled in the precision trap.
So, in the case of wrong capture times, the ions have larger
axial amplitudes, come closer to the holes
in the endcaps and thus are more influenced by the correction tubes.

\section{Summary and conclusion}
In order to achieve accurate mass values employing Penning trap mass
spectrometry the electric and magnetic fields in the Penning trap have to
be optimized to correct for deviations arising from geometrical
trap imperfections and homogeneity limits of the superconducting
magnet. The procedure for the optimization of the electric and
magnetic field for the precision Penning trap of ISOLTRAP has been
described and demonstrated. The optimization was performed by
varying the capture time, i.e. the delay period between the ejection
of iond from the preparation trap and the capturing in the precision trap,
thus giving them different axial energy, i.e. changing the amplitude of
their axial oscillation. The magnetic field was optimized first to
avoid any influence on the electric field optimization. 

\section*{Acknowledgments}
This work was supported by the German Federal Ministry for Education
and Research (BMBF) under contract no. 06GF151 and 06MZ215, by the
Helmholtz Association of National Research Centers (HGF) under
contract no. VH-NG-037, by the European Commission under contract
no. HPRI-CT-2001-50034 (NIPNET), by the Marie Curie fellowship
network HPMT-CT-2000-00197, and by the French IN2P3.


\begin{thebibliography}{00}
\bibitem{Lun03} D. Lunney, J.M. Pearson, and C. Thibault, Rev. Mod. Phys. 75 (2003) 1021.
\bibitem{Bla06} K. Blaum \textit{et al.}, Phys. Rep. 425 (2006) 1.
\bibitem{Rai04} S. Rainville, J.K. Thompson, and D.E. Pritchard, Science 303 (2004) 334.
\bibitem{Bol96} G. Bollen \textit{et al.}, Nucl. Instrum. and Meth. A 368 (1996) 675.
\bibitem{Muk08} M. Mukherjee \textit{et al.}, Eur. Phys. J. A 35 (2008) 1.
\bibitem{Kug00} E. Kugler, Hyperfine Interact. 129 (2000) 23.
\bibitem{Kel03} A. Kellerbauer \textit{et al.}, Eur. Phys. J. D 22 (2003) 53.
\bibitem{Her01} F. Herfurth \textit{et al.}, Nucl. Instrum. and Meth. A 469 (2001) 254.
\bibitem{Her05} A. Herlert \textit{et al.}, New J. Phys. 7 (2005) 44.
\bibitem{Bla02} K. Blaum \textit{et al.}, Eur. Phys. J. A 15 (2002) 245.
\bibitem{Bla03b} K. Blaum \textit{et al.}, Anal. Bioanal. Chem. 377 (2003) 1133.
\bibitem{Bla03} K. Blaum \textit{et al.}, J. Phys. B 36 (2003) 921.
\bibitem{Geo2007a} S. George \textit{et al.}, Phys. Rev. Lett. 98 (2007) 162501.
\bibitem{Geo2007b} S. George \textit{et al.}, Int. J. Mass Spectrom. 264 (2007) 110.
\bibitem{Sch06} Special issue on Ultra-Accurate Mass Spectrometry
and Related Topics, edited by L. Schweikhard and G. Bollen [Int. J.
Mass Spectrom. 251 (2006) 85].
\bibitem{Ber02} I. Bergstr\"om \textit{et al.}, Nucl. Instrum. and Meth. A 487 (2002) 618.
\bibitem{Sav06} G. Savard \textit{et al.}, Int. J. Mass Spectrom. 251 (2006) 252.
\bibitem{Jok06} A. Jokinen \textit{et al.}, Int. J. Mass Spectrom. 251 (2006) 204.
\bibitem{Rah06} S. Rahaman \textit{et al.}, Int. J. Mass Spectrom. 251 (2006) 146.
\bibitem{Rin06} R. Ringle \textit{et al.}, Int. J. Mass Spectrom. 251 (2006) 300.
\bibitem{Dil06} J. Dilling \textit{et al.}, Int. J. Mass Spectrom. 251 (2006) 198.
\bibitem{Hab05} D. Habs \textit{et al.}, Eur. Phys. J. A 25 S01 (2005) 57.
\bibitem{Ket08} J. Ketelaer \textit{et al.}, Nucl. Instrum. and Meth. A, submitted (2008).
\bibitem{Bro86} L.S. Brown and G. Gabrielse, Rev. Mod. Phys. 58 (1986) 233.
\bibitem{Bol90} G. Bollen \textit{et al.}, J. Appl. Phys. 68 (1990) 4355.
\bibitem{Kre90} M. Kretzschmar, Phys. Scripta 46 (1992) 545 and 555.
\bibitem{Bro82} L.S. Brown and G. Gabrielse, Phys. Rev. A 25 (1982) 2423.
\bibitem{Yaz06} C. Yazidjian \textit{et al.}, Hyperfine Interact. 173 (2006) 181.
\bibitem{Sav91} G. Savard \textit{et al.}, Phys. Lett. A 158 (1991) 247.
\bibitem{Rai97} H. Raimbault-Hartmann \textit{et al.}, Nucl. Instrum. Meth. B 126 (1997) 378.
\bibitem{Gra80} G. Gr\"aff, H. Kalinowsky and J. Traut, Z. Phys. A 297 (1980) 35.
\bibitem{Konig} M. K\"onig \textit{et al.}, Int. J. Mass Spectrom. Ion Process. 142 (1995) 95.
\bibitem{MMJ07} M. Marie-Jeanne \textit{et al.}, Nucl. Instrum. and Meth. A 587 (2008) 464.
\bibitem{Ger90} Ch. Gerz, D. Wildsorf, and G. Werth, Nucl. Instrum. and Meth. B 47 (1990) 453.
\end{thebibliography}
\end{document}